\documentclass[letterpaper,preprintnumbers,prd,twocolumn,nofootinbib,nobibnotes,showpacs]{revtex4}
\usepackage{amsfonts}
\usepackage{mathrsfs}
\usepackage{epsfig}
\usepackage{graphicx}%
\usepackage{dcolumn}
\usepackage{amsmath}
\usepackage{color}
\usepackage{color}
\makeatletter
\def\btt#1{\texttt{\@backslashchar#1}}%
\DeclareRobustCommand\bblash{\btt{\@backslashchar}}%
\makeatother
\begin{document}

\title{Quintessence and phantom emerging from the split-complex field and the split-quaternion field}
\author{Changjun Gao}\email{gaocj@bao.ac.cn} \affiliation{Key Laboratory for Computational Astrophysics, National Astronomical
Observatories, Chinese Academy of Sciences, Beijing, 100012,
China} \affiliation{State Key Laboratory of Theoretical Physics,
Institute of Theoretical Physics, Chinese Academy of Sciences,
Beijing 100190, China}
\author{Xuelei Chen}\email{xuelei@cosmology.bao.ac.cn} \affiliation{Key Laboratory for Computational Astrophysics, National Astronomical
Observatories, Chinese Academy of Sciences, Beijing, 100012,
China} \affiliation{Center of High Energy Physics, Peking
University, Beijing 100871, China}
\author{You-Gen Shen}
\email{ygshen@center.shao.ac.cn} \affiliation{Shanghai
Astronomical Observatory, Chinese Academy of Sciences, Shanghai
200030, China}

\date{\today}

\begin{abstract}

Motivated by the mathematic theory of split-complex numbers (or
hyperbolic numbers, also perplex numbers) and the split-quaternion
numbers (or coquaternion numbers), we define the notion of
split-complex scalar field and the split-quaternion scalar field.
Then we explore the cosmic evolution of these scalar fields in the
background of spatially flat Friedmann-Robertson-Walker Universe.
We find that both the quintessence field and the phantom field
could naturally emerge in these scalar fields. Introducing the
metric of field space, these theories fall into a subclass of the
multi-field theories which have been extensively studied in
inflationary cosmology.

\end{abstract}

\pacs{98.80.Jk, 04.40.Nr, 04.50.+h, 11.25.Mj}

\maketitle

\section{Introduction}
More and more accurate and convincing astronomical observations
\cite{wmap:13,am:10,blake:11} indicate that dark energy dominates
our current Universe. Although a host of observationally viable
dark energy models have been proposed, the nature of dark energy
is still undetermined. The Einstein cosmological constant is in
many respects the most economical solution to the problem of dark
energy. But it is confronted with two fundamental problems: the
fine tuning problem and the coincidence problem (see e.g.
\cite{wein:00}). So one turn to the study of other options for
dark energy, for example, quintessence
\cite{ratra:88,wet:88,lid:98}, quintom \cite{quintom:04},
k-essence \cite{kessence:99}, Chaplygin gas \cite{chap:01},
holographic dark energy \cite{holo:04} and so on.

One find that in many models of quintessence field, there exist
the so-called tracker solutions. In these solutions, the
quintessence field always has a energy density closely tracks (but
is smaller than) that of the radiation until the matter-radiation
equality. By this way, the coincidence problem is solved
\cite{ratra:88,wet:88,lid:98}. Phantom energy is introduced into
the study of cosmic evolution by Caldwell ~\cite{cal:02,cal:03}.
The Lagrangian of phantom takes the form
$\mathscr{L}_{\phi}=-\frac{1}{2}\nabla_{\mu}\phi\nabla^{\mu}\phi+V\left(\phi\right)$,
with $\phi$ the phantom field and $V(\phi)$ the phantom potential.
By changing the sign of kinetic term in quintessence field
\emph{by hand,} this form of Lagrangian could be obtained. One
then find that the energy density of phantom actually increases
with cosmic time. So the fate of the Universe is a Big Rip.

Now we want to ask: can we unify quintessence and phantom in a
natural way? The answer is yes. We find that the quintessence and
phantom can naturally emerge in the theory of split-complex scalar
field, the split-quaternion scalar field and the split-octonion
scalar field. This finding is motivated by the mathematic theory
of split-complex numbers (or hyperbolic numbers, also perplex
numbers), the split-quaternion numbers (or coquaternion numbers)
and the split-octonion numbers \cite{ant:93} (for references, see
also the Wikipedia
\footnote{{{$http://en.wikipedia.org/wiki/Split-complex \
number.$}}}).

We note that our application of split-complex number
is not just a mathematical tool but rather it has physical information too. Actually, the theory of split-complex number
has been applied in gravitational fields \cite{kun:83}, quantum group
\cite{wei:97}, quantum mechanics \cite{bra:03}, quintessence cosmology \cite{wei:03} and so on.
What is more, with the help of split-complex variables, a pseudo-complex field theory \cite{hess:07} and
a pseudo-complex General Relativity \cite{hess:08} are recently presented. For the split-complex scalar field, if one require it obeys the symmetry of invariance under hyperbolic rotation, the split-complex field would restore to the Hessence field
\cite{wei:05}. On the other hand, if the symmetry is allowed to be broken, the split-complex scalar field would turn out to be the quintom field \cite{quintom:04}.
Coleman found that
for conventional complex field, there exists $\textrm{Q-Balls}$ solutions \cite{col:85} due to the conserved charge.
We find there is also conserved charge for split-complex field. So we expect $\textrm{Q-Balls}$-like solutions exist which makes the split-complex field
more physical. Finally, we find the split-complex scalar field could be re-formulated as one specific model of the multi-filed theory which attracts a lot of effort in the study of inflationary universe \cite{ko:00,ko:01,ko:02,lang:00,lang:01} in recent years. In all, the split scalar fields have rich physics.

The paper is organized as follows. In section~\ref{sec:split}, we
define the notion of split-complex scalar field and show
quintessence and phantom could emerge from this field. In
section~\ref{sec:splitdy}, we investigate the cosmic evolution of
the split-complex field and show that the detail of dynamics is
closely related to the initial conditions on the quintessence and
phantom. In section~\ref{sec:qua}, we investigate the cosmic
evolution of the split-quaternion field. In
section~\ref{sec:stable}, the linear perturbations of these fields
in the background of Friedmann-Robertson-Walker Universe are
present. Conclusions and discussions are given in
section~\ref{sec:conclusion}. Throughout this paper, we adopt the
system of units in which $G=c=\hbar=1$ and the metric signature
$(-,\ +,\ +,\ +)$.

\section{split-complex scalar field}\label{sec:split}
\subsection{What is split-complex scalar field}

The mathematic theory of split-complex numbers (or hyperbolic
numbers, also perplex numbers) can be found in Ref.~\cite{ant:93}
or the Wikipedia
\footnote{{{$http://en.wikipedia.org/wiki/Split-complex \
number.$}}}. Motivated by the theory of these numbers, we define
the split-complex scalar field ${{\Phi}}$ as follows
\begin{eqnarray}\label{eq:phi}
\Phi=\phi+j\psi\;,
\end{eqnarray}
where $\phi$ and $\psi$ are two real scalar fields. The quantity
$j$ is similar to the imaginary unit $i$ except that \cite{ant:93}
\begin{eqnarray}
j^2=+1\;.
\end{eqnarray}
Choosing $j^2=-1$ results in the conventional complex scalar
field. It is this change of sign which distinguishes the
split-complex scalar field from the ordinary complex one. The
quantity $j$ here is not a real number but an independent
quantity. Namely, it is not equal to $\pm 1$.

Just as for ordinary complex field, one can define the notion of a
split-complex conjugate as follows
\begin{eqnarray}
\Phi^{*}=\phi-j\psi\;.
\end{eqnarray}
Then the modulus of a split-complex scalar field is given by the
isotropic quadratic form
\begin{eqnarray}
\Phi\Phi^{*}=\phi^2-\psi^2\;.
\end{eqnarray}
This quadratic form is split into positive and negative parts, in
contrast to the positive definite form of the ordinary complex
scalar field.

Similar to the ordinary complex field which can be written in the
form of Euler's formula
\begin{eqnarray}
\Phi=\phi e^{i\theta}=\phi\cos\theta+i\phi\sin\theta\;,
\end{eqnarray}
the split-complex field has the Euler's formula as follows
\begin{eqnarray}
\Phi=\phi e^{j\theta}=\phi\cosh\theta+j\phi\sinh\theta\;.
\end{eqnarray}
This can be derived from a power series expansion using the fact
that $cosh$ has only even powers while that for $sinh$ has odd
powers. It follows that $\Phi\Phi^{*}=\phi^2$.
\subsection{Quintessence and phantom from split-complex field}
We shall consider the theory of a massive, split-complex,
self-interacting scalar field with the Lagrangian density as
follows

\begin{eqnarray}\label{eq:L}
\mathscr{L}=\frac{1}{2}\nabla_{\mu}\Phi\nabla^{\mu}\Phi^{*}+\frac{1}{4}\lambda^2\left(\Phi\Phi^{*}+m^2/\lambda^2\right)^2\;,
\end{eqnarray}
with $m$ the mass of the scalar and $\lambda$ a coupling constant. Here the potential is the same as minimally coupled
Higgs potential which has been given experimental evidence. Of course, one may consider other potentials, i.e, exponential potential,
power-law potential and so on. However, if one demand the theory obeys the symmetry of hyperbolic rotation (see subsection $\textrm{C}$ below),
the potential should be constrained to be the function of $\Phi\Phi^{*}$, namely $V(\Phi\Phi^{*})$. On the other hand,
if the symmetry is broken, the potential could be $V(\Phi+\Phi^{*},\ \Phi-\Phi^{*})$. In the next, we will see the former is exactly the
Hessence field \cite{wei:05} and the latter the quintom field \cite{quintom:04}. So the split-complex scalar field procedure
is not just a mathematical re-formulation of the Hessence or Quintom, but has the physical meaning of internal symmetry.

\begin{figure}[h]
\begin{center}
\includegraphics[width=9cm]{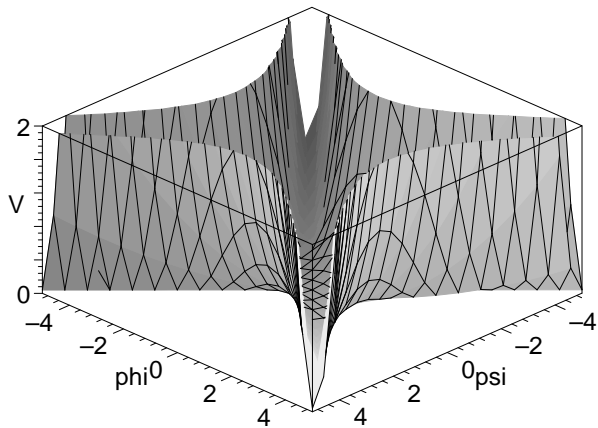}

\caption{The sketch of the potential
$V\propto(\Phi\Phi^{*}+M^2)^2$ with the quintessence field $\phi$
and the phantom field $\psi$. The potential has the vanishing
absolute vacuum energy on the hyperbola $\phi^2-\psi^2+M^2=0$.}.
\label{pot}
\end{center}
\end{figure}

Substituting Eq.~(\ref{eq:phi}) into Eq.~(\ref{eq:L}), we obtain

\begin{eqnarray}\label{eq:hessence}
\mathscr{L}&=&\frac{1}{2}\nabla_{\mu}\phi\nabla^{\mu}\phi-\frac{1}{2}\nabla_{\mu}\psi\nabla^{\mu}\psi
\nonumber\\&&+\frac{1}{4}\lambda^2\left(\phi^2-\psi^2+m^2/\lambda^2\right)^2\;.
\end{eqnarray}
If we take the field $\Phi$ in the Lagrangian Eq.~(\ref{eq:L}) as
the ordinary complex scalar field, $\Phi=\phi+i\psi$, the
Lagrangian takes the form
\begin{eqnarray}\label{eq:ordinary}
\mathscr{L}&=&\frac{1}{2}\nabla_{\mu}\phi\nabla^{\mu}\phi+\frac{1}{2}\nabla_{\mu}\psi\nabla^{\mu}\psi
\nonumber\\&&+\frac{1}{4}\lambda^2\left(\phi^2+\psi^2+m^2/\lambda^2\right)^2\;.
\end{eqnarray}
It is apparent there is sign of difference before the terms
$\frac{1}{2}\nabla_{\mu}\psi\nabla^{\mu}\psi$ and $\psi^2$. This
is due to the fact that $i^2=-1$ and $j^2=1$. One can recognize
that Lagrangian, Eq.~(\ref{eq:hessence}) is nothing but the
Hessence field proposed by Wei et al in Ref.~\cite{wei:05}. $\phi$
and $\psi$ plays the role of quintessence and phantom,
respectively. The difference of Hessence field from the quintom
field \cite{quintom:04}
\begin{eqnarray}\label{eq:quin}
\mathscr{L}&=&\frac{1}{2}\nabla_{\mu}\phi\nabla^{\mu}\phi-\frac{1}{2}\nabla_{\mu}\psi\nabla^{\mu}\psi+V\left(\phi,
\psi\right)\;,
\end{eqnarray}
is that the form of the scalar potential is greatly constrained by
$V(\phi^2-\psi^2)$ in Hessence. We point out that there is a
remarkable difference in our motivation from the Hessence.
Ref.~\cite{wei:05} propose the Lagrangian density of Hessence as
follows
\begin{eqnarray}\label{eq:lh}
\mathscr{L}&=&\frac{1}{2}\nabla_{\mu}\Phi\nabla^{\mu}\Phi+\frac{1}{2}\nabla_{\mu}\Phi^{*}\nabla^{\mu}\Phi^{*}+V\left(\Phi^2+
\Phi^{*2}\right)\;,
\end{eqnarray}
with $\Phi$ the conventional complex scalar field. However, our
Lagrangian density Eq.~(\ref{eq:L}) is different from
Eq.~(\ref{eq:lh}) not only on the Lagrangian expression but also
on the physical meaning of scalar field $\Phi$. In
Fig.~{\ref{pot}} we plot the scalar potential
$V\propto(\Phi\Phi^{*}+M^2)^2$ with the quintessence field $\phi$
and the phantom field $\psi$. The potential has the vanishing
absolute vacuum energy on the hyperbola $\phi^2-\psi^2+M^2=0$.
\subsection{symmetry}
The theory of Lagrangian Eq.~(\ref{eq:L}) has the symmetry that it
is invariant after a hyperbolic rotation
\begin{eqnarray}
\Phi\rightarrow \Phi e^{j\alpha}\;,
\end{eqnarray}
with $\alpha$ a constant. Expressed the split-complex scalar field
in terms of two real fields $\phi$ and $\psi$, the hyperbolic
rotation corresponds to the $O(1,1)$ transformations
\begin{eqnarray}
&&\phi\rightarrow \phi \cosh\alpha+\psi\sinh\alpha\;,\\&&
\psi\rightarrow \phi \sinh\alpha+\psi\cosh\alpha\;.
\end{eqnarray}
Then the No$\ddot{e}$ther theorem tells us this symmetry leads to
a conserved charge which is given by the formula
\begin{eqnarray}
Q=\frac{1}{2j}\int
d^3x\left(\Phi^{*}\partial_0\Phi-\Phi\partial_0\Phi^{*}\right)\;,
\end{eqnarray}
in the background of four dimensional Minkowski spacetime. Here
$\partial_0$ represents the derivative with the time. Coleman found that
for conventional complex field, there would exist $\textrm{Q-Balls}$ solutions \cite{col:85} due to the conserved charge.
Since there is also conserved charge for split-complex field, we expect $\textrm{Q-Balls}$-like solutions also exist which makes the split-complex field
more physical.

\subsection{the stability}
Introduce the metric tensor $\eta_{IJ}$ in the field space
$\Phi_I=(\phi, \psi)$,
\begin{eqnarray}\label{eq:m0}
\eta_{IJ}=\eta^{IJ}=\left(%
\begin{array}{cc}\label{eq:eta}
  1 & 0 \\
  0 & -1 \\
\end{array}%
\right)\;,
\end{eqnarray}
where $I,J=1,2$. Eq.~(\ref{eq:hessence}) can be expressed as

\begin{eqnarray}\label{eq:m1}
\mathscr{L}=\frac{1}{2}\nabla_{\mu}\Phi_I\nabla^{\mu}\Phi^{I}+\frac{1}{4}\lambda^2\left(\Phi_I\Phi^{I}+m^2/\lambda^2\right)^2\;.
\end{eqnarray}
With this form, the theory belongs to the general multiple scalar
field theory considered in
Ref.~\cite{ko:00,ko:01,ko:02,lang:00,lang:01}. The equation of
motion and the energy momentum are give by
\begin{eqnarray}
\nabla_{\mu}\nabla^{\mu}{\Phi_I}-\lambda^2\left(\Phi_J\Phi^{J}+m^2/\lambda^2\right)\Phi_I=0\;,
\end{eqnarray}
and
\begin{eqnarray}
T_{\mu\nu}=-\nabla_{\mu}\Phi_I\nabla^{\mu}{\Phi^I}+g_{\mu\nu}\mathscr{L}\;.
\end{eqnarray}
From the equation of motion, we see both the quintessence
$\phi$ and the phantom $\psi$ acquire a vanishing effective mass
$m_0$:
\begin{eqnarray}
m_0=\lambda\sqrt{\Phi_J\Phi^{J}+m^2/\lambda^2}|_{\left(\Phi_J\Phi^{J}+m^2/\lambda^2\right)}=0\;,
\end{eqnarray}
at the global minimum of the potential.
In this case, not only quintessence but also phantom is stable in dynamics.

This is different from the usual phantom field with the
Lagrangian
\begin{eqnarray}
\mathscr{L}=-\frac{1}{2}\nabla_{\mu}\psi\nabla^{\mu}{\psi}+V\left(\psi\right)\;,
\end{eqnarray}
and the equation of motion
\begin{eqnarray}
\nabla_{\mu}\nabla^{\mu}{\psi}+m_0^2\psi=0\;,
\end{eqnarray}
where $m_0^2\equiv V_{,\psi\psi}\mid_{\psi=\psi_0}$. The value of
$\psi=\psi_0$ corresponds to the global minimum of the potential.
In this case, the phantom field acquires an \emph{imaginary }mass
which would lead to the classically and quantum instability. Then
why is there such a difference in the mass? The reason are as
follows. The usual phantom scalar potential has the form
$V=\frac{1}{2}m_0^2\psi^2$ at the global minimum. However, for the
Lagrangian Eq.~(\ref{eq:hessence}), the phantom potential takes
the form $V=-\frac{1}{2}m_0^2\psi^2$ at the global minimum. Thus it
is very important that the potential of the split-complex scalar
is constrained to be $V(\phi^2-\psi^2)$ in order to avoid the
problem of instability .

\subsection{conformal invariant split-complex field}
Recently, Kallosh and Linde \cite{KL:13} proposed a simple
conformally invariant two-field model of $\textrm{dS/AdS}$ space.
The model consists of two real scalar fields, $\phi$ and $\psi$,
which has the $SO(1, 1)$ symmetry:
\begin{eqnarray}
\mathscr{L}_{KL}&=&\frac{1}{2}\nabla_{\mu}\phi\nabla^{\mu}\phi-\frac{1}{2}\nabla_{\mu}\psi\nabla^{\mu}\psi+\frac{1}{12}\left(\phi^2-\psi^2\right)R
\nonumber\\&&-\frac{1}{4}\lambda\left(\phi^{2}-\psi^2\right)^2\;.
\end{eqnarray}
Here $R$ is the Ricci scalar and $\lambda$ a coupling constant.
This theory is locally conformal invariant under the following
transformations,
\begin{eqnarray}
\tilde{g}_{\mu\nu}=e^{-2\sigma\left(x\right)}g_{\mu\nu}\;,\ \ \
\tilde{\phi}=e^{\sigma\left(x\right)}\phi\;,\ \
\tilde{\psi}=e^{\sigma\left(x\right)}\psi\;.
\end{eqnarray}
The global $SO(1, 1)$ symmetry is a boost between these two
fields. Using the concept of our split-complex scalar field
$\Phi_I$, the theory is equivalent to
\begin{eqnarray}
\mathscr{L}=\frac{1}{2}\nabla_{\mu}\Phi_I\nabla^{\mu}\Phi^{I}+\frac{1}{12}R\Phi_I\Phi^I
-\frac{1}{4}\lambda\left(\Phi_I\Phi^{I}\right)^2\;.
\end{eqnarray}
In other words, the two scalar fields considered in
Ref.~\cite{KL:13} is exactly the conformal invariant split-complex
field.

\section{dynamics of split-complex scalar field}\label{sec:splitdy}
In this section, we investigate the cosmic evolution of the
split-complex field in the background of spatially flat
Friedmann-Robertson-Walker (FRW) Universe
\begin{eqnarray}
ds^2=-dt^2+a\left(t\right)^2\left(dr^2+r^2d\Omega^2\right)\;,
\end{eqnarray}
where $a(t)$ is the cosmic scale factor. We model all other matter
sources present in the Universe as perfect fluids. These matter
sources can be baryonic matter, relativistic matter and dark
energy. We assume there is no interaction between the
split-complex scalar field and other matter fields, other than by
gravity. Then the Einstein equations and the equation of motion of
the scalar fields are given by
\begin{eqnarray}
3H^2&=&\kappa^2\left(\frac{1}{2}\dot{\phi}^2-\frac{1}{2}\dot{\psi}^2+V+\rho_r+\rho_m\right)\;,\\
2\dot{H}+3H^2&=&-\kappa^2\left(\frac{1}{2}\dot{\phi}^2-\frac{1}{2}\dot{\psi}^2-V+\frac{1}{3}\rho_r\right)\;,
\end{eqnarray}
and
\begin{eqnarray}
\ddot{\phi}+3H\dot{\phi}+V_{,\phi}&=&0\;,\\
\ddot{\psi}+3H\dot{\psi}-V_{,\psi}&=&0\;,
\end{eqnarray}
respectively. Here $H\equiv\dot{a}/a$ denotes the Hubble parameter
and dot is the derivative with respect to cosmic time, $t$.
$\rho_m$ and $\rho_r$ are the energy density of dark matter and
relativistic matter. $V_{,\phi}$ and $V_{,\psi}$ denote the
derivative with respect to $\phi$ and $\psi$, respectively.
Introduce the following dimensionless quantities
\begin{eqnarray}
&&X\equiv\frac{\kappa}{\sqrt{6}}\frac{\dot{\phi}}{H}\;,\  \  \  \
Y\equiv\frac{\kappa}{\sqrt{6}}\frac{\dot{\psi}}{H}\;,\nonumber\\&&
Z\equiv\frac{\kappa}{\sqrt{6}}\frac{\sqrt{V}}{H}\;,\  \  \
U\equiv\frac{\kappa}{\sqrt{3}}\frac{\sqrt{\rho_m}}{H}\;,\nonumber\\&&
\lambda_{\phi}\equiv-\frac{2}{\sqrt{6}}\frac{V_{,\phi}}{\kappa
V}\;,\  \
\lambda_{\psi}\equiv-\frac{2}{\sqrt{6}}\frac{V_{,\psi}}{\kappa
V}\;,\ \nonumber\\ &&  N\equiv\ln a\;,
\end{eqnarray}
then the equations of motion can be written in the following
autonomous form
\begin{eqnarray}
\frac{dX}{dN}&=&-3X-\frac{3}{2}\lambda_{\phi}Z^2-X\frac{\dot{H}}{H^2}\;,\nonumber\\
\frac{dY}{dN}&=&-3Y-\frac{3}{2}\lambda_{\psi}Z^2-Y\frac{\dot{H}}{H^2}\;,\nonumber\\
\frac{dZ}{dN}&=&-\frac{3}{2}Z\left(\lambda_{\phi}X+\lambda_{\psi}Y\right)-Z\frac{\dot{H}}{H^2}\;,\nonumber\\
\frac{dU}{dN}&=&-\frac{3}{2}U-U\frac{\dot{H}}{H^2}\;,\nonumber\\
\frac{d\lambda_{\phi}}{dN}&=&\frac{3X\left(\lambda_{\psi}^2-\lambda_{\phi}^2\right)}{2+2\sqrt{1-s^2\left(\lambda_{\phi}^2-\lambda_{\psi}^2\right)}}
\nonumber\\&&+\frac{3}{2}\lambda_{\phi}\left(X\lambda_{\phi}-Y\lambda_{\psi}\right)\;,\nonumber\\
\frac{d\lambda_{\psi}}{dN}&=&\frac{3Y\left(\lambda_{\phi}^2-\lambda_{\psi}^2\right)}{2+2\sqrt{1-s^2\left(\lambda_{\phi}^2-\lambda_{\psi}^2\right)}}
\nonumber\\&&+\frac{3}{2}\lambda_{\psi}\left(X\lambda_{\phi}-Y\lambda_{\psi}\right)\;,
\end{eqnarray}
together with a constraint equation
\begin{eqnarray}
X^2-Y^2+Z^2+U^2+\frac{\kappa^2\rho_r}{3H^2}=1\;.
\end{eqnarray}
Here
\begin{eqnarray}
s=\frac{\sqrt{6}}{4}\frac{m\kappa}{\lambda}\;,
\end{eqnarray}
and
\begin{eqnarray}
\frac{\dot{H}}{H^2}=-2-X^2+Y^2+2Z^2+\frac{1}{2}U^2\;.
\end{eqnarray}

The equation of state $w$ and the fraction of the energy density
for the split-complex scalar field are

\begin{eqnarray}
w&\equiv&\frac{X^2-Y^2-Z^2}{X^2-Y^2+Z^2}\;,\nonumber\\
\Omega_{\Phi}&\equiv& X^2-Y^2+Z^2\;,
\end{eqnarray}

As an example, we consider $s=1$. Physically, the quintessence
$\phi$ would roll down the potential and the phantom $\psi$ roll
up the potential. Therefore we can safely assume $X>0, Y> 0,
\lambda_{\phi}>0,\lambda_{\psi}<0$. We investigate the cosmology
model with the following values of parameters: $\Omega_{k0}=0,
\Omega_{m0}=0.27, \Omega_{r0}=8.1 \cdot 10^{-5},
\Omega_{X0}=0.73$, which are consistent with current observations
\cite{spergel:2007}.

\begin{figure}[h]
\begin{center}
\includegraphics[width=9cm]{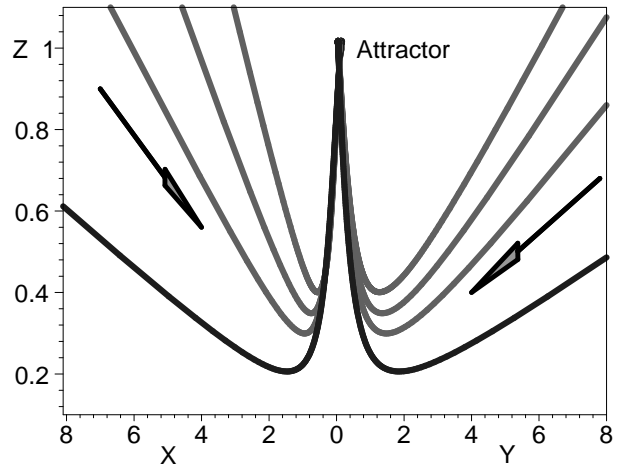}
\caption{The phase plane for the evolution of the split-complex
scalar with a range of different initial conditions. The point (0,
0, 0) corresponds to the radiation or matter dominated epoches.
The point (0, 0, 0) is unstable and the point (0, 0, 1) is stable
and thus an attractor. These trajectories show that the Universe
always ends at the split-complex scalar potential dominated
epoch.} \label{phase0}
\end{center}
\end{figure}

In Fig.~\ref{phase0}, we plot the phase plane for the evolution of
the split-complex scalar with a range of different initial
conditions. The point (0, 0, 0) corresponds to the radiation or
matter dominated epoches. The point (0, 0, 0) is unstable and the
point (0, 0, 1) is stable and thus an attractor. These
trajectories show that the Universe always ends at the
split-complex scalar potential dominated epoch.

\begin{figure}[h]
\begin{center}
\includegraphics[width=9cm]{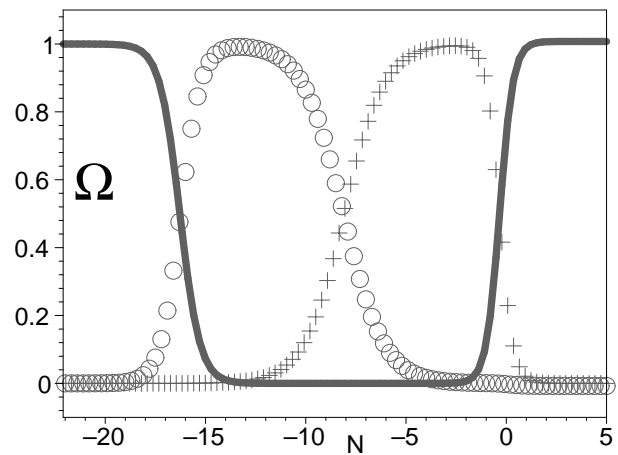}

\caption{The evolution of density fractions for radiation (circled
line), dark matter (crossed line) and the split-complex scalar
field(solid line)} \label{phase}
\end{center}
\end{figure}

In Fig.~\ref{phase}, we plot the evolution of density fractions
for radiation, dark matter and the split-complex scalar field.
This shows that the split-complex scalar field can mimic the dark
energy very well.

\begin{figure}[h]
\begin{center}
\includegraphics[width=9cm]{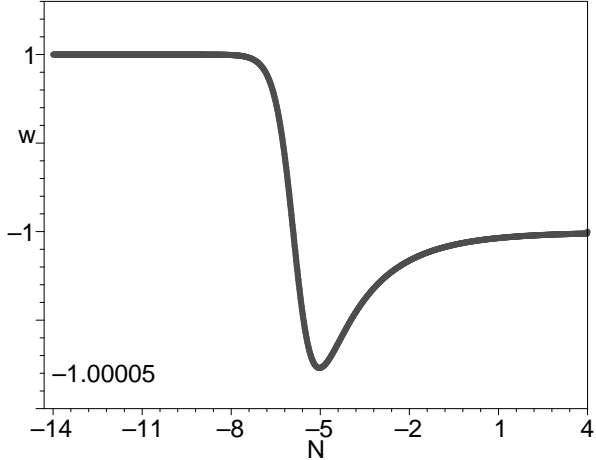}

\caption{The evolution of the equation of state for the
split-complex scalar. It behaves as the stiff matter at higher
redshifts and a cosmological constant for the lower redshifts. The
split-complex scalar can cross the phantom divide around the value
of $N=-5$ (redshift $3000$)}. \label{wx}
\end{center}
\end{figure}

In Fig.~\ref{wx}, we plot the evolution of the equation of state
for the split-complex scalar. It behaves as the stiff matter at
higher redshifts and a cosmological constant for the lower
redshifts. It can cross the phantom divide around the value of
$N=-5$ (redshift $3000$).


\section{dynamics of split-quaternion scalar field}\label{sec:qua}
\subsection{ split-quaternion scalar field}
In this subsection, we define the notion of split-quaternion
scalar field and the split-octonion field. We find they actually
consists of two quintessence fields and two phantom fields. The
number of quintessence is exactly the same as that of the phantom.
To this end, we start from the theory of split-quaternion number
$q$ (or coquaternion) which is given by
\cite{jame:49}\footnote{{{$http://en.wikipedia.org/wiki/Split-quaternion.$}}}
\begin{eqnarray}
q=u+ix+jy+kz\;,
\end{eqnarray}
where $u,x,y,z$ are real numbers. The quantity $i,j,k$ here are
not real numbers but independent quantities. The products of these
elements are \cite{jame:49}
\begin{eqnarray}
&&ij=k =-ji,\nonumber\\&& jk = -i =-kj,\nonumber\\&& ki
=j=-ik,\nonumber\\&& i^2 =-1,\nonumber\\&& j^2 =+1,\nonumber\\&&
k^2=+1,
\end{eqnarray}
and hence $ijk=1$. A split-quaternion has a conjugate
\begin{eqnarray}
q^{*}=u-ix-jy-kz\;,
\end{eqnarray}
and multiplicative modulus
\begin{eqnarray}
qq*=u^2 + x^2-y^2-z^2.
\end{eqnarray}
This quadratic form is split into positive and negative parts, in
contrast to the positive definite form on the algebra of
quaternions.

We now define the split-quaternion scalar field $\Phi$ as
\begin{eqnarray}
\Phi=\phi_1+i\phi_2+j\psi_1+k\psi_2.
\end{eqnarray}
Then the Lagrangian
\begin{eqnarray}\label{eq:sq}
\mathscr{L}=\frac{1}{2}\nabla_{\mu}\Phi\nabla^{\mu}\Phi^{*}+\frac{1}{4}\lambda^2\left(\Phi\Phi^{*}+\frac{6\Lambda}{\kappa^2}\right)^2\;,
\end{eqnarray}
can be written as
\begin{eqnarray}\label{eq:sqq}
\mathscr{L}&=&\frac{1}{2}\nabla_{\mu}\phi_1\nabla^{\mu}\phi_1+\frac{1}{2}\nabla_{\mu}\phi_2\nabla^{\mu}\phi_2\nonumber\\&&-
\frac{1}{2}\nabla_{\mu}\psi_1\nabla^{\mu}\psi_1-\frac{1}{2}\nabla_{\mu}\psi_2\nabla^{\mu}\psi_2\nonumber\\&&+
\frac{1}{4}\lambda^2\left(\phi_1^2+\phi_2^2-\psi_1^2-\psi_2^2+\frac{6\Lambda}{\kappa^2}\right)^2\;.
\end{eqnarray}
Here $\Lambda$ is a positive constant. It is apparent the theory
consists of two quintessence field $\phi_1, \phi_2$ and two
phantom field $\psi_1,\psi_2$. Introduce the metric tensor
$\eta_{IJ}$ in the space of field
$\Phi_I=(\phi_1,\phi_2,\psi_1,\psi_2)$,
\begin{eqnarray}
\eta_{IJ}=\eta^{IJ}=\left(%
\begin{array}{cccc}
  1& 0 &  0  & 0\\
  0 & 1 &  0  & 0\\
  0 & 0 &  -1 & 0\\
  0 & 0 &  0  & -1\\
\end{array}%
\right)\;,
\end{eqnarray}
where $I,J=1,2,3,4$. Eq.~(\ref{eq:sqq}) can be expressed as
\begin{eqnarray}\label{eq:m2}
\mathscr{L}=\frac{1}{2}\nabla_{\mu}\Phi_I\nabla^{\mu}\Phi^{I}+\frac{1}{4}
\lambda^2\left(\Phi_I\Phi^{I}+\frac{6\Lambda}{\kappa^2}\right)^2\;.
\end{eqnarray}
With this form, the theory also belongs to the general multiple
scalar field theory considered in
Ref.~\cite{ko:00,ko:01,ko:02,lang:00,lang:01}. Furthermore, we
could define the split-octonion field (motivated by the
split-octonion\footnote{$http://en.wikipedia.org/wiki/Split-octonion.$})
by introducing the metric tensor in the space of field
$\Phi_I=(\phi_1,\phi_2,\phi_3,\phi_4,\psi_1,\psi_2,\psi_3,\psi_4)$,

\begin{eqnarray}
\eta_{IJ}=\eta^{IJ}=\left(%
\begin{array}{cccccccc}
  1 & 0 &  0  & 0 & 0 & 0 & 0 & 0\\
   0 & 1 &  0  & 0 & 0 & 0 & 0 & 0\\
    0 & 0 &  1  & 0 & 0 & 0 & 0 & 0\\
     0 & 0 &  0  & 1 & 0 & 0 & 0 & 0\\
      0 & 0 &  0  & 0 & -1 & 0 & 0 & 0\\
    0 & 0 &  0  & 0 & 0 & -1 & 0 & 0\\
   0 & 0 &  0  & 0 & 0 & 0 & -1 & 0\\
  0 & 0 &  0  & 0 & 0 & 0 & 0 & -1\\
         \end{array}%
\right)\;,
\end{eqnarray}
where $I,J=1,2,3,4,5,6,7,8$. The Lagrangian Eq.~(\ref{eq:m2}) can
also describe a massive, self-interacting split-octonion scalar
field.

\subsection{dynamics}
In this section, we study the dynamics of the split-quaternion
scalar field in the background of spatially flat FRW Universe. For
simplicity, we only consider the cosmic evolution of the
split-quaternion scalar field.

The Einstein equations and the equation of motion of the scalar
fields take the form

\begin{eqnarray}
&&3H^2=\kappa^2\left(\frac{1}{2}\dot{\Phi}_I\dot{\Phi}^I+V\right)\;,\nonumber\\
&&2\dot{H}+3H^2=-\kappa^2\left(\frac{1}{2}\dot{\Phi}_I\dot{\Phi}^I-V\right)\;,
\end{eqnarray}
and
\begin{eqnarray}
\ddot{\Phi}_I+3H\dot{\Phi}_I+V_{,\Phi^{I}}&=&0\;,
\end{eqnarray}
respectively.

Introduce the following dimensionless quantities
\begin{eqnarray}
&&X\equiv\frac{\kappa}{\sqrt{6}}\frac{\dot{\phi}_1}{H}\;,\  \
\lambda_{\phi_1}\equiv\frac{3}{\sqrt{6}\kappa}\frac{{V_{,\phi_1}}}{V}\;,\nonumber\\&&
Y\equiv\frac{\kappa}{\sqrt{6}}\frac{\dot{\phi}_2}{H}\;,\ \
\lambda_{\phi_2}\equiv\frac{3}{\sqrt{6}\kappa}\frac{{V_{,\phi_2}}}{V}\;,\nonumber\\&&
Z\equiv\frac{\kappa}{\sqrt{6}}\frac{\dot{\psi}_1}{H}\;,\ \
\lambda_{\psi_1}\equiv-\frac{3}{\sqrt{6}\kappa}\frac{{V_{,\psi_1}}}{V}\;,\nonumber\\&&
U\equiv\frac{\kappa}{\sqrt{6}}\frac{\dot{\psi}_2}{H}\;,\ \
\lambda_{\psi_2}\equiv-\frac{3}{\sqrt{6}\kappa}\frac{{V_{,\psi_2}}}{V}\;,\nonumber\\&&
\nonumber\\ &&  N\equiv\ln a\;,
\end{eqnarray}
then the equations of motion can be written in the following
autonomous form
\begin{eqnarray}
\frac{dX}{dN}&=&-3X-XB-\lambda_{\phi_1}A\;,\nonumber\\
\frac{dY}{dN}&=&-3Y-YB-\lambda_{\phi_2}A\;,\nonumber\\
\frac{dZ}{dN}&=&-3Z-ZB-\lambda_{\psi_1}A\;,\nonumber\\
\frac{dU}{dN}&=&-3U-UB-\lambda_{\psi_2}A\;,\nonumber\\
\frac{d\lambda_{\phi_1}}{dN}&=&-X\lambda_{\phi_1}^2-Y\lambda_{\phi_1}\lambda_{\phi_2}
-Z\lambda_{\phi_1}\lambda_{\psi_1}-U\lambda_{\phi_1}\lambda_{\psi_2}\nonumber\\&&+\frac{XC}{1+\sqrt{1-\Lambda
C}}\;,\nonumber\\
\frac{d\lambda_{\phi_2}}{dN}&=&-X\lambda_{\phi_1}\lambda_{\phi_2}-Y\lambda_{\phi_2}^2
-Z\lambda_{\phi_2}\lambda_{\psi_1}-U\lambda_{\phi_2}\lambda_{\psi_2}\nonumber\\&&+\frac{YC}{1+\sqrt{1-\Lambda
C}}\;,\nonumber\\
\frac{d\lambda_{\psi_1}}{dN}&=&X\lambda_{\phi_1}\lambda_{\psi_1}+Y\lambda_{\phi_2}\lambda_{\psi_1}
+Z\lambda_{\psi_1}^2+U\lambda_{\psi_2}\lambda_{\psi_1}\nonumber\\&&+\frac{ZC}{1+\sqrt{1-\Lambda
C}}\;,\nonumber\\
\frac{d\lambda_{\psi_2}}{dN}&=&X\lambda_{\phi_1}\lambda_{\psi_2}+Y\lambda_{\phi_2}\lambda_{\psi_2}
+Z\lambda_{\psi_1}\lambda_{\psi_2}+U\lambda_{\psi_2}^2\nonumber\\&&+\frac{UC}{1+\sqrt{1-\Lambda
C}}\;,\nonumber\\
A&\equiv& 1-X^2-Y^2+Z^2+U^2\;,\nonumber\\
B&\equiv& -3\left(X^2+Y^2-Z^2-U^2\right)\;, \nonumber\\
C&\equiv&
\lambda_{\phi_1}^2+\lambda_{\phi_2}^2-\lambda_{\psi_1}^2-\lambda_{\psi_2}^2\;,
\end{eqnarray}
together with a constraint equation
\begin{eqnarray}
X^2+Y^2-Z^2-U^2+\frac{\kappa^2 V}{3H^2}=1\;.
\end{eqnarray}

The equation of state $w$ of the split-quaternion scalar field is

\begin{eqnarray}
w&\equiv&-1-\frac{2}{3}\frac{\dot{H}}{H^2}=-1-\frac{2}{3}B\;.
\end{eqnarray}

We note that the coupling constant $\lambda$ is not present in
above equations. So there is only one free parameter, $\Lambda$.
As an example, we consider $\Lambda=1$. Physically, the
quintessence $\phi_1, \phi_2$ would roll down the potential and
the phantom $\psi_1,\psi_2$ roll up the potential. Therefore we
shall assume
$\lambda_{\phi_1}>0,\lambda_{\phi_2}>0,\lambda_{\psi_1}<0,\lambda_{\psi_2}<0,
X<0, Y<0, Z>0, U>0$.

\begin{figure}[h]
\begin{center}
\includegraphics[width=9cm]{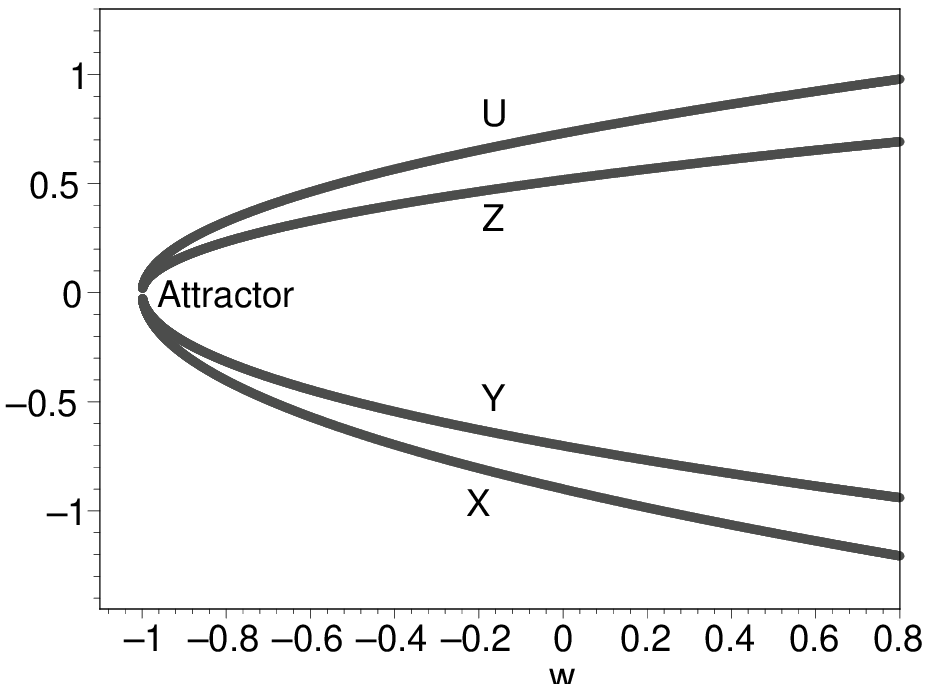}

\caption{Evolution of the scaled velocities $X, Y, Z, U$ for the
quintessence fields $\phi_1$, $\phi_2$ and the phantom fields
$\psi_1$, $\psi_2$, respectively, with respect to the equation of
state $w$. The initial values are put by $X(0)=-1.22,\
Y(0)=-0.95,\ Z(0)=0.99,\ U(0)=0.70,\ \lambda_{\phi_1}(0)=1/20,\
\lambda_{\phi_2}(0)=1/20,\ \lambda_{\psi_1}(0)=-1/20,\
\lambda_{\psi_2}(0)=-1/20$. In other words, we set the same
strength of fields but different scaled velocities. Here and in
the afterwards,
$(\cdot\cdot\cdot)(0)\equiv(\cdot\cdot\cdot)(N=0)$. The figure
shows that if the kinetic energy of quintessence dominates over
that of phantom initially (with the same strength of fields),
$X^2+Y^2>Z^2+U^2$, the equation of state of the split-quaternion
field would evolve from $+1$ to $-1$ with the point
$(X,Y,Z,U,w)=(0,0,0,0,-1)$ as the attractor.}. \label{f1}
\end{center}
\end{figure}

\begin{figure}[h]
\begin{center}
\includegraphics[width=9cm]{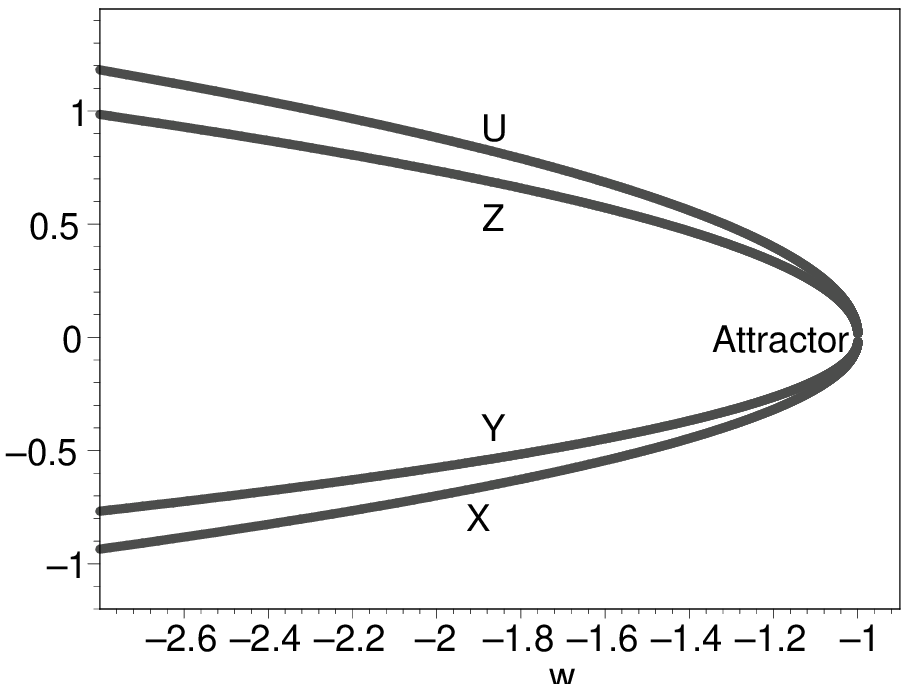}

\caption{Evolution of the scaled velocities $X, Y, Z, U$ for the
quintessence fields $\phi_1$, $\phi_2$, the phantom fields
$\psi_1$, $\psi_2$, respectively, with respect to the equation of
state $w$. The initial values are put by $X(0)=-0.95,\
Y(0)=-0.78,\ Z(0)=1.2,\ U(0)=1.0,\ \lambda_{\phi_1}(0)=1/20,\
\lambda_{\phi_2}(0)=1/20,\ \lambda_{\psi_1}(0)=-1/20,\
\lambda_{\psi_2}(0)=-1/20$. In other words, we set the same
strength of the fields but different scaled velocities. The figure
shows that if the kinetic energy of phantom dominates over that of
quintessence initially, $X^2+Y^2<Z^2+U^2$ (with the same strength
of fields), the equation of state of the split-quaternion field
would evolve from minus infinity to $-1$ with the point
$(X,Y,Z,U,w)=(0,0,0,0,-1)$ as the attractor.} \label{f2}
\end{center}
\end{figure}

\begin{figure}[h]
\begin{center}
\includegraphics[width=9cm]{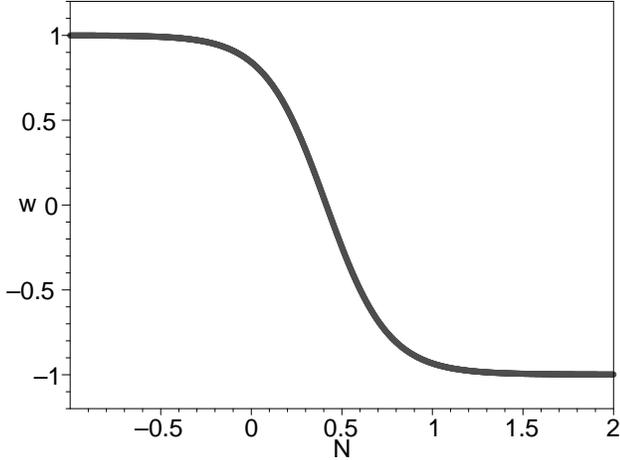}

\caption{The evolution of the equation of state for the
split-quaternion scalar when the kinetic energy of quintessence
dominates over that of phantom (with the same strength of fields).
It behaves as the stiff matter at higher redshifts and a
cosmological constant for the lower redshifts.}. \label{e1}
\end{center}
\end{figure}

\begin{figure}[h]
\begin{center}
\includegraphics[width=9cm]{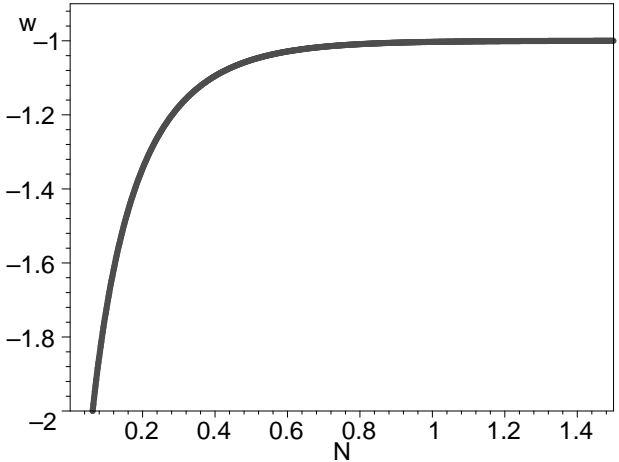}

\caption{The evolution of the equation of state for the
split-quaternion scalar when the kinetic energy of phantom
dominates over that of quintessence (with the same strength of
fields). The equation of state is always smaller than unit one.}.
\label{e2}
\end{center}
\end{figure}

In Fig.~\ref{f1} and Fig.~\ref{f2}, we plot the evolution of the
scaled velocities $X,Y,Z,U$ for the quintessence fields $\phi_1$,
$\phi_2$ and the phantom fields $\psi_1$, $\psi_2$, respectively,
with respect to the equation of state $w$. They show that if the
kinetic energy of quintessence dominates over that of phantom
initially, $X^2+Y^2>Z^2+U^2$ ($X(0)=-1.22,\ Y(0)=-0.95,\
Z(0)=0.99,\ U(0)=0.70,\ \lambda_{\phi_1}(0)=1/20,\
\lambda_{\phi_2}(0)=1/20,\ \lambda_{\psi_1}(0)=-1/20,\
\lambda_{\psi_2}(0)=-1/20$), the equation of state of the
split-quaternion field would evolve from $+1$ to $-1$ with the
point $(X,Y,Z,U,w)=(0,0,0,0,-1)$ as the attractor (Fig.~\ref{f1}).
On the other hand,  if the kinetic energy of phantom dominates
over that of quintessence initially, $X^2+Y^2<Z^2+U^2$
($X(0)=-0.95,\ Y(0)=-0.78,\ Z(0)=1.2,\ U(0)=1.0,\
\lambda_{\phi_1}(0)=1/20,\ \lambda_{\phi_2}(0)=1/20,\
\lambda_{\psi_1}(0)=-1/20,\ \lambda_{\psi_2}(0)=-1/20$), the
equation of state of the split-quaternion field would evolve from
minus infinity to $-1$ with the point $(X,Y,Z,U,w)=(0,0,0,0,-1)$
as the attractor (Fig.~\ref{f2}). Corresponding to Fig.~\ref{f1}
and Fig.~\ref{f2}, we plot the equation of state in Fig.~\ref{e1}
and Fig.~\ref{e2}, respectively.

\begin{figure}[h]
\begin{center}
\includegraphics[width=9cm]{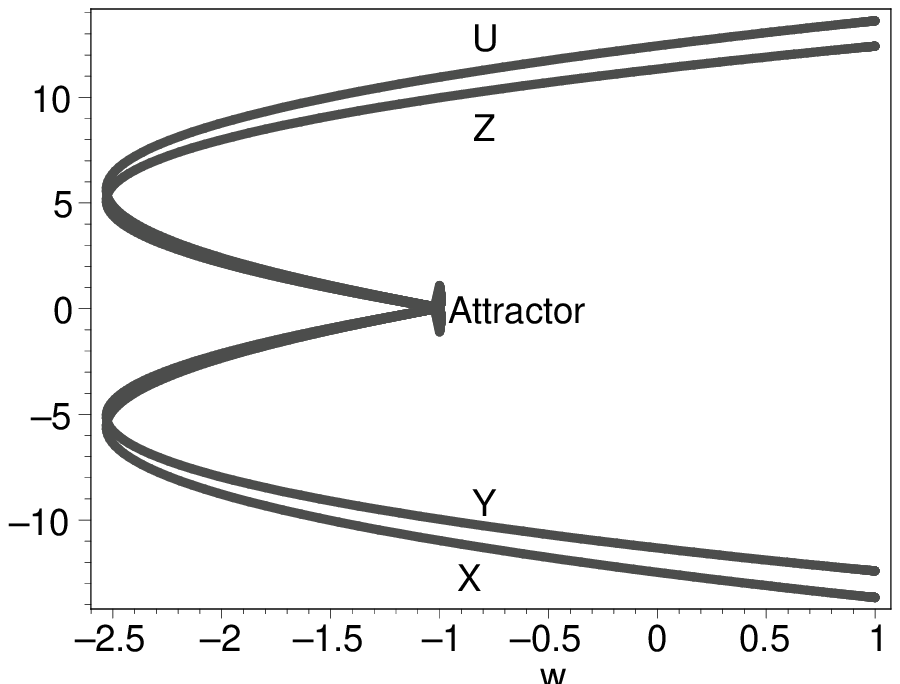}

\caption{Evolution of the scaled velocities $X, Y, Z, U$ for the
quintessence fields $\phi_1$, $\phi_2$, the phantom fields
$\psi_1$, $\psi_2$, respectively, with respect to the equation of
state $w$. The initial values are put by $X(0)=-1,\ Y(0)=-1.1,\
Z(0)=1,\ U(0)=1.1,\ \lambda_{\phi_1}(0)=1/20,\
\lambda_{\phi_2}(0)=1/20,\ \lambda_{\psi_1}(0)=-1/27,\
\lambda_{\psi_2}(0)=-1/20$. The figure shows that the equation of
state of the split-quaternion field could cross the phantom divide
with the point $(X,Y,Z,U,w)=(0,0,0,0,-1)$ as the attractor.}.
\label{f3}
\end{center}
\end{figure}

In Fig.~\ref{f3}, we plot the evolution of the scaled velocities
$X, Y, Z, U$ for the quintessence fields $\phi_1$, $\phi_2$, the
phantom fields $\psi_1$, $\psi_2$, respectively, with respect to
the equation of state $w$. The initial values are put by
$X(0)=-1,\ Y(0)=-1.1,\ Z(0)=1,\ U(0)=1.1,\
\lambda_{\phi_1}(0)=1/20,\ \lambda_{\phi_2}(0)=1/20,\
\lambda_{\psi_1}(0)=-1/27,\ \lambda_{\psi_2}(0)=-1/20$. The figure
shows that the equation of state of the split-quaternion field
could cross the phantom divide with the point
$(X,Y,Z,U,w)=(0,0,0,0,-1)$ as the attractor.

\begin{figure}[h]
\begin{center}
\includegraphics[width=9cm]{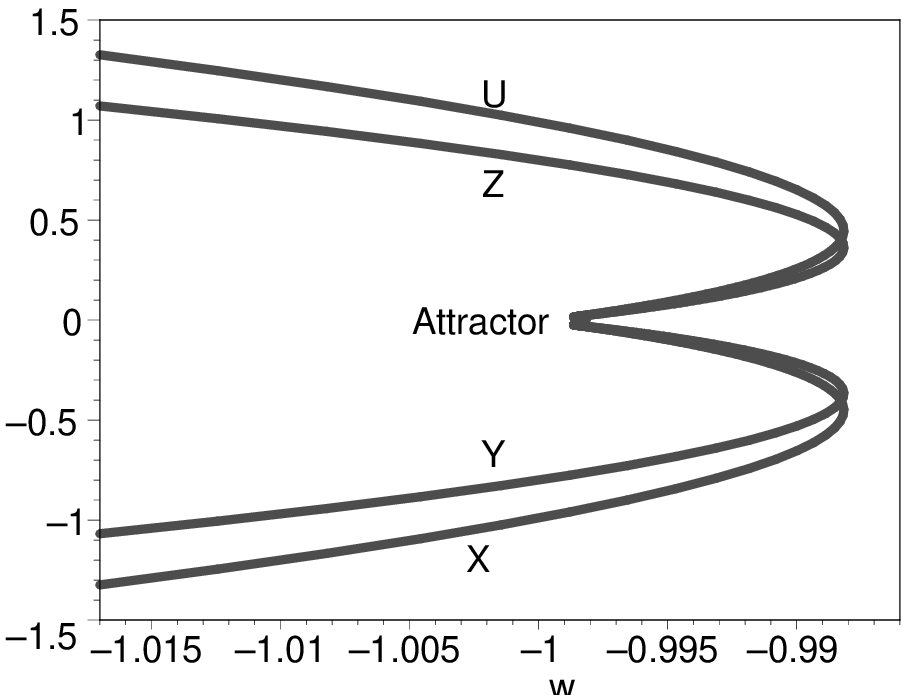}

\caption{Evolution of the scaled velocities $X, Y, Z, U$ for the
quintessence fields $\phi_1$, $\phi_2$, the phantom fields
$\psi_1$, $\psi_2$, respectively, with respect to the equation of
state $w$. The initial values are put by $X(0)=-0.8,\ Y(0)=-0.99,\
Z(0)=0.805,\ U(0)=0.99,\ \lambda_{\phi_1}(0)=1/15,\
\lambda_{\phi_2}(0)=1/15,\ \lambda_{\psi_1}(0)=-1/20,\
\lambda_{\psi_2}(0)=-1/20$. The figure shows that the equation of
state of the split-quaternion field cold cross the phantom divide
with the point $(X,Y,Z,U,w)=(0,0,0,0,-1)$ as the attractor.}.
\label{f4}
\end{center}
\end{figure}

\begin{figure}[h]
\begin{center}
\includegraphics[width=9cm]{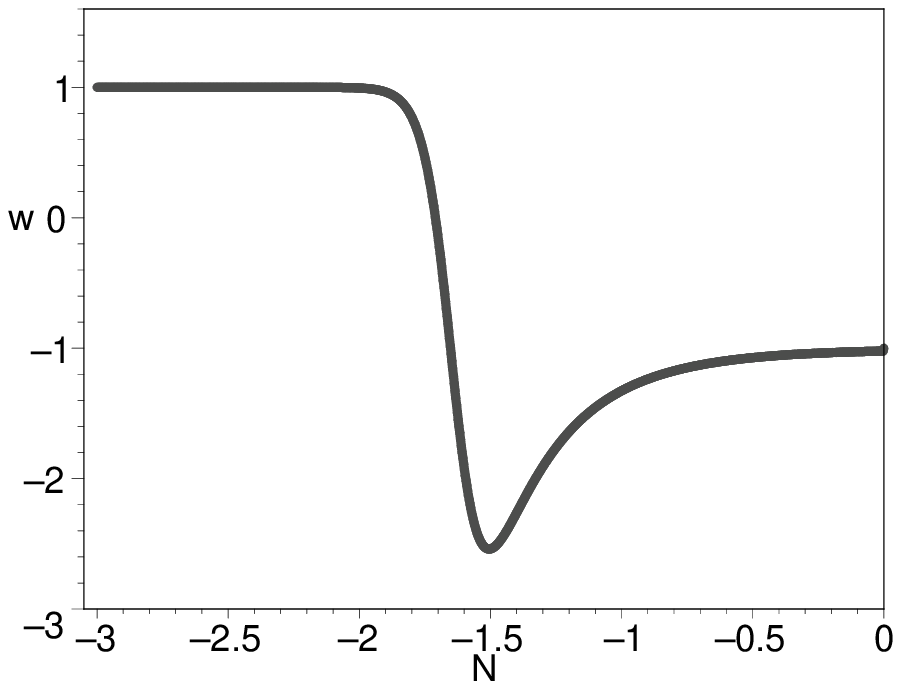}

\caption{The evolution of the equation of state for the
split-complex scalar corresponding Fig.~(\ref{f3}). It behaves as
the stiff matter at higher redshifts and a cosmological constant
for the lower redshifts. The split-complex scalar can cross the
phantom divide around the value of $N=-1.5$ }. \label{e3}
\end{center}
\end{figure}

\begin{figure}[h]
\begin{center}
\includegraphics[width=9cm]{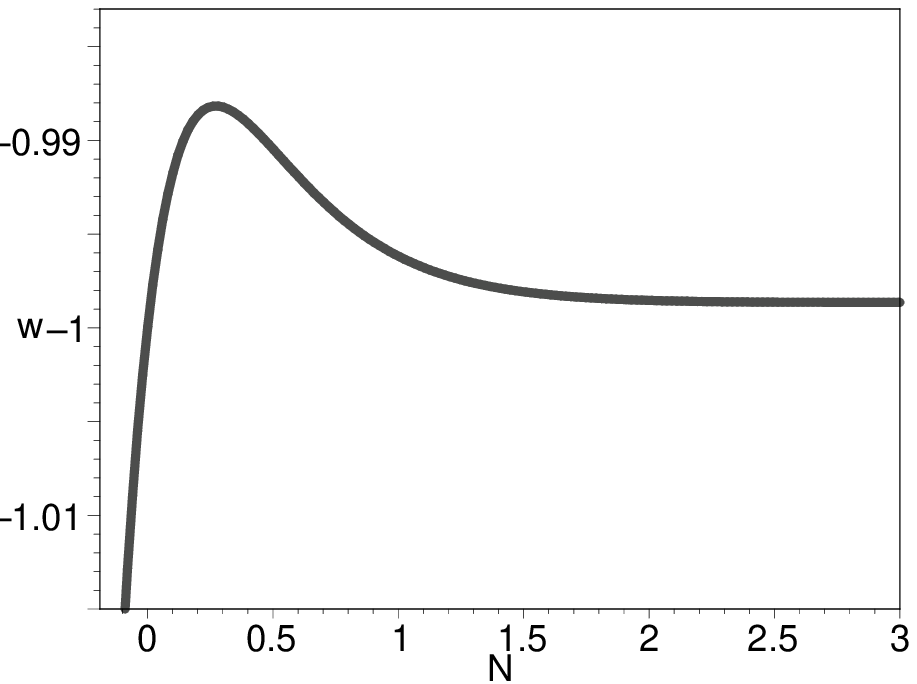}

\caption{The evolution of the equation of state for the
split-complex scalar corresponding Fig.~(\ref{f4}). It behaves as
the phantom matter at higher redshifts and a cosmological constant
for the lower redshifts. The split-complex scalar can cross the
phantom divide around the value of $N=0.3$}. \label{ef}
\end{center}
\end{figure}

In Fig.~\ref{f4}, we plot evolution of the scaled velocities $X,
Y, Z, U$ for the quintessence fields $\phi_1$, $\phi_2$, the
phantom fields $\psi_1$, $\psi_2$, respectively, with respect to
the equation of state $w$. The initial values are put by
$X(0)=-0.8,\ Y(0)=-0.99,\ Z(0)=0.805,\ U(0)=0.99,\
\lambda_{\phi_1}(0)=1/15,\ \lambda_{\phi_2}(0)=1/15,\
\lambda_{\psi_1}(0)=-1/20,\ \lambda_{\psi_2}(0)=-1/20$. The figure
shows that the equation of state of the split-quaternion field
cold cross the phantom divide with the point
$(X,Y,Z,U,w)=(0,0,0,0,-1)$ as the attractor. In Fig.~(\ref{e3})
and Fig.~(\ref{ef}), we plot the evolution of their equation of
state, respectively.

In all, the detail of dynamics of the split-quaternion scalar
field is closely related to the initial conditions on the fields.

\subsection{multi-field theory}
In the proceeding sections, we have studied the cosmic evolution
of the spilt-complex field and the split-quaternion field with the
Lagrangian as follows,
$\mathscr{L}=\frac{1}{2}\partial_{\mu}\Phi^{I}\partial^{\mu}\Phi_{I}+V$,
Eq.~(\ref{eq:m1}) and (\ref{eq:m2}). In this subsection, we extend
the expression of Lagrangian and show that they actually belong to
the multi-field theories studied in the inflationary cosmology
\cite{ko:00,ko:01,ko:02,lang:00,lang:01}. To this end, we define

\begin{eqnarray}\label{eq:X}
X^{IJ}\equiv-\frac{1}{2}\partial_{\mu}\Phi^{I}\partial^{\mu}\Phi^{J}\;.
\end{eqnarray}
In the spirit of k-inflation \cite{armen:99}, the very general
Lagrangian is of the form
\begin{eqnarray}
\mathscr{L}=K\left(X^{IJ},\ \Phi^{I}\right)\;,
\end{eqnarray}
where $I= 1,\cdot\cdot\cdot,N$ labels the multiple fields. Here
$\Phi_I$ can be split-complex field ($N=2$), split-quaternion
field (N=4), split-octonion field (N=8) and so on. The field
indices $I,J$ are raised and lowered with the metric tensor
$\eta_{IJ}$ and $\eta^{IJ}$ of the field space. If $\eta^{IJ}$ is
replaced with the most general metric $\tilde{g}_{IJ}(\Phi^{K})$,
it is just the extensively studied multi-field theory in
inflationary cosmology \cite{ko:00,ko:01,ko:02,lang:00,lang:01}.

\section{linear perturbations}\label{sec:stable}
Up to now, we have explored the cosmic evolution of split-complex
field and the split-quaternion field, respectively. In this
section, we focus on the linear perturbations for the
split-complex field. The extension of the result to
split-quaternion and split-octonion field is straightforward.

We start from the action as follows (following the convention and
notation in Ref.~\cite{ko:00})
\begin{equation}
S=\int d^4x\sqrt{-g}\tilde{P}(Y, \Phi^I)\,,
\end{equation}
where
\begin{equation}
Y=\eta_{IJ}X^{IJ} + \frac{b\left(\Phi^K\right)}{2}
\left(X^2-X^{J}_I X^I_J\right)\,,
\end{equation}
where $\Phi^K$ is the split-complex field. $\eta_{IJ}$ is defined
by Eq.~(\ref{eq:eta}). When $b=0$, the Lagrangian,
Eq.~(\ref{eq:m1}) is included as a particular case of
$\tilde{P}(X,\ \Phi^K)$.

In order to study the evolution of linear perturbations in the
background of FRW Universe, we expand the above action to second
order, including both metric and scalar field perturbations. In
the uniform curvature gauge, the perturbed split-complex field
takes the form
\begin{eqnarray}
\Phi^I\left(x,t\right)=\Phi_0^I\left(t\right)+Q^I\left(x,t\right),
\label{deltaphigauge}
\end{eqnarray}
where $Q^I$ denotes the field perturbations. In the following, we
will usually drop the subscript $``0"$ on $\Phi_0^I$ and simply
identify $\Phi^I$ as the fields in FRW Universe unless otherwise
stated.

The second order action can be expressed as
\begin{eqnarray}
S_{(2)} &=& \frac12 \int dt d^3 x a^3 \biggl[ \left(
\tilde{P}_{,Y}  \eta_{IJ} + \tilde{P}_{,YY} \dot{\Phi}_I
\dot{\Phi}_J \right) \dot{Q}^I \dot{Q}^J
\nonumber\\
&&- \frac{1}{a^2}  \tilde{P}_{,Y} \left[(1+bX) \eta_{IJ} - b
X_{IJ}\right]
\partial_i Q^I \partial^i Q^J
\nonumber\\&&-\bar{{\cal{M}}}_{IJ} Q^I Q^J + 2 \tilde{P}_{,YJ}
\dot{\Phi}_I Q^J \dot{Q}^I
\biggr]\,,\nonumber\\
\end{eqnarray}
with the effective squared mass matrix
\begin{eqnarray}
\bar{{\cal{M}}}_{IJ} &=&-\tilde{P}_{,IJ}+ \frac{X
\tilde{P}_{,Y}}{H} (\tilde{P}_{,YJ} \dot{\Phi}_I
+ \tilde{P}_{,YI} \dot{\Phi}_J)\nonumber\\
&&+\frac{X \tilde{P}^3 _{,Y}}{2 H^2}(1-\frac{1}{c_{ad}^2})
\dot{\Phi}_I \dot{\Phi}_J\nonumber\\&&- \frac{1}{a^3}
\left[\frac{a^3}{2H} \tilde{P}^2 _{,Y}
\left(1+\frac{1}{c_{ad}^2}\right)\dot{\Phi}_I \dot{\Phi}_J
\right]^{\cdot}\,.
\end{eqnarray}
Here $\tilde{P}_{,Y}$ and $\tilde{P}_{,J}$ denote the derivative
of $\tilde{P}$ with respect to $Y$ and $\Phi^J$. Dot denotes the
derivative with respect to cosmic time $t$. $a$ and $H$ are the
scale factor and Hubble parameter of the Universe, respectively.
$c_{ad}$ is the sound speed for adiabatic perturbations defined by
\begin{equation}
c_{ad}^{2} \equiv \frac{\tilde{P}_{,Y}}{\tilde{P}_{,Y} + 2X
\tilde{P}_{,YY}}\;.
\end{equation}
The sound speed squared of entropy perturbations is defined by
\begin{equation}
\quad c_{en}^{2} \equiv 1 + bX\,.
\end{equation}

For Lagrangian, Eq.~(\ref{eq:m1}), we have
\begin{equation}
 c_{ad}=c_{en}=1\,,
\end{equation}
and the second order action is simply
\begin{eqnarray}
S_{(2)} &=& \frac12 \int dt d^3 x a^3 \biggl[\dot{Q}_I \dot{Q}^I
-\frac{1}{a^2}\partial_i Q_I \partial^i Q^I
-\bar{{\cal{M}}}_{IJ} Q^I Q^J
\biggr]\,,\nonumber\\
\end{eqnarray}
where
\begin{eqnarray}
\bar{{\cal{M}}}_{IJ} &=&\lambda^2\Phi_I\Phi_J- \frac{1}{a^3}
\left(\frac{a^3}{H}\dot{\Phi}_I \dot{\Phi}_J
\right)^{\cdot}\,.
\end{eqnarray}
The equation of motion is given by
\begin{eqnarray}
\ddot{Q}_I-\frac{1}{a^2}\partial_i \partial^i Q_I+\bar{{\cal{M}}}_{IJ}Q^{J}=0\,.
\end{eqnarray}
In the gauge of $Q^2=0$, we have
\begin{eqnarray}
\ddot{Q}^1-\frac{1}{a^2}\partial_i \partial^i Q^1+\lambda^2\phi^2 Q^{1}=0\,.
\end{eqnarray}
The squared mass term is positive. Thus the perturbation to quintessence $Q^{1}$ is stable.
On the other hand, in the gauge of $Q^1=0$, we have
\begin{eqnarray}
\ddot{Q}^2-\frac{1}{a^2}\partial_i \partial^i Q^2-\lambda^2\psi^2 Q^{2}=0\,.
\end{eqnarray}
In this case, the squared mass term is negative and the perturbation to phantom $Q^{2}$ is unstable.

In order to achieve a scale invariant power spectrum, we now decompose the perturbations into adiabatic and entropy
part as follows, $Q^I = Q_\sigma e^I_\sigma + Q_s e^I_s$, where
\begin{equation}
e^I_\sigma=e_1^I = \frac{\dot{\Phi}^I}{\sqrt{P_{,X^{JK}}
\dot{\Phi}^J \dot{\Phi}^K}}\,,
\end{equation}
and $e^I_s$ is the unit vector orthogonal to $e^I_\sigma$. Use the
conformal time $\tau = \int dt /a(t)$ and define the canonically
normalized fields
\begin{eqnarray}
v_\sigma \equiv \frac{a}{c_{ad}} Q_\sigma\,,\;\;\;\;\; v_s \equiv
\frac{a}{c_{en}}Q_s\,,
\end{eqnarray}
one derive the equations of motion for $v_\sigma$ and $v_s$
\begin{eqnarray}
&&v_s''+\xi v_\sigma'+\left(c_{en}^2 k^2-\frac{\alpha''}{\alpha} +
a^2 \mu_s^2\right)v_s-\frac{z'}{z}\xi v_\sigma = 0\;,\nonumber
\\&& v_\sigma '' - \xi v_s' + \left( c_{ad}^2 k^2 -\frac{z''}{z}
\right) v_\sigma - \frac{(z \xi)'}{z}v_s =0\,.
\end{eqnarray}
Here the prime denotes the derivative with respect to $\tau$ and
\begin{eqnarray}
\xi &\equiv& \frac{a}{\dot{\sigma} \tilde{P}_{,Y} c_{ad}} \left[
(1+ c_{ad}^2) \tilde{P}_{,s} -c_{ad}^2 \dot{\sigma}^2
\tilde{P}_{,Ys}\right]\,,\label{coupling} \nonumber\\  \mu_s^2
&\equiv& -\frac{\tilde{P}_{,ss}}{\tilde{P}_{,Y}}- \frac{1}{2
c_{ad}^2 X} \frac{\tilde{P}_{,s}^2}{\tilde{P}_{,Y}^2} +2
\frac{\tilde{P}_{,Ys} \tilde{P}_{,s}}{\tilde{P}_{,Y}^2}\,,
\nonumber\\ z &\equiv& \frac{a \dot{\sigma}}{ c_{ad} H}
\sqrt{\tilde{P}_{,Y}}\,, \ \ \  \;\;\;\;\;\alpha\equiv a
\sqrt{\tilde{P}_{,Y}}\,,
\end{eqnarray}
with
\begin{eqnarray}
\dot{\sigma} &\equiv& \sqrt{2X}\,,\;\;\;\;\; \ \ \  \tilde{P}_{,s}
\equiv
\tilde{P}_{,I} e^I_s \sqrt{\tilde{P}_{,Y}} c_{en}\,,\nonumber\\
\;\;\;\;\;\tilde{P}_{,Ys} &\equiv& \tilde{P}_{,YI} e^I_s
\sqrt{\tilde{P}_{,Y}} c_{en}\,,\nonumber\\
\;\;\;\;\;\tilde{P}_{,ss} &\equiv& \tilde{P}_{,IJ} e^I_s e^J_s
\tilde{P}_{,Y} c_{en}^2 \,.
\end{eqnarray}
In the limit of weak coupling, $\xi\simeq 0$, small effective
mass, $\mu_s\simeq 0$ and slow-roll, $\dot{H}\simeq 0$, one have
$z^{''}/z=2/\tau^2$ and $\alpha^{''}/\alpha=2/\tau^2$. So the
equations of motion turn out to be
\begin{eqnarray}
&&v_s''+\left(c_{en}^2 k^2-\frac{2}{\tau^2} \right)v_s=
0\;,\nonumber
\\&& v_\sigma '' + \left( c_{ad}^2 k^2 -\frac{2}{\tau^2}
\right) v_\sigma =0\,.
\end{eqnarray}
Refs.~\cite{wands:99} and ~\cite{finelli:02} have ever shown that
above equations lead to exact scale invariant power spectrum.

\section{conclusion and discussion}\label{sec:conclusion}

In conclusion, motivated by the mathematic theory of split-complex
numbers, the split-quaternion numbers and split-octonion numbers,
we have proposed the notion of split-complex scalar field,
split-quaternion scalar field and the split-octonion scalar field.
We note that our proposal of split scalar fields is not just a mathematical tool but rather it has physical information.

In the first place, the well-known quintessence and phantom fields could
naturally emerge in these fields. So in order to construct a
phantom field, one need not resort to quintessence by changing the
sign of its kinetic term \emph{by hand.} On the other hand, the
conventional complex scalar field usually can not generate a
phantom field by the decomposition, $\Phi=\phi+i\psi$ except for a
non-canonical form of Lagrangian \cite{wei:05}. Secondly, if one require the split-complex scalar field obeys the symmetry of invariance under hyperbolic rotation, the split-complex field
would restore to the Hessence field \cite{wei:05}. But if the symmetry of invariance is allowed to be broken, the split-complex scalar
 field would turn out to be the quintom field \cite{quintom:04}. So compared to Hessence and quintom, they have rich physics. Thirdly, we find there is a conserved charge for the split-complex field. Therefore, the Coleman's $\textrm{Q-Balls}$-like solutions are expected to exist in the split-complex field which makes the field even more physical.
 Finally, by introducing the metric of field space, these split scalar fields fall into a
subclass of the multi-field theories which are being studied in inflationary cosmology \cite{ko:00,ko:01,ko:02,lang:00,lang:01}. Also, following the usual procure of quantization of the complex scalar field \cite{quan:15}, one could carry out the quantization of split-complex scalar field.

\acknowledgments

This work is supported by the Chinese MoST
863 program under grant 2012AA121701, the NSFC under grant
11373030, 10973014, 11373020 and 11465012.

\newcommand\ARNPS[3]{~Ann. Rev. Nucl. Part. Sci.{\bf ~#1}, #2~ (#3)}
\newcommand\AL[3]{~Astron. Lett.{\bf ~#1}, #2~ (#3)}
\newcommand\AP[3]{~Astropart. Phys.{\bf ~#1}, #2~ (#3)}
\newcommand\AJ[3]{~Astron. J.{\bf ~#1}, #2~(#3)}
\newcommand\APJ[3]{~Astrophys. J.{\bf ~#1}, #2~ (#3)}
\newcommand\APJL[3]{~Astrophys. J. Lett. {\bf ~#1}, L#2~(#3)}
\newcommand\APJS[3]{~Astrophys. J. Suppl. Ser.{\bf ~#1}, #2~(#3)}
\newcommand\JHEP[3]{~JHEP.{\bf ~#1}, #2~(#3)}
\newcommand\JCAP[3]{~JCAP. {\bf ~#1}, #2~ (#3)}
\newcommand\LRR[3]{~Living Rev. Relativity. {\bf ~#1}, #2~ (#3)}
\newcommand\MNRAS[3]{~Mon. Not. R. Astron. Soc.{\bf ~#1}, #2~(#3)}
\newcommand\MNRASL[3]{~Mon. Not. R. Astron. Soc.{\bf ~#1}, L#2~(#3)}
\newcommand\NPB[3]{~Nucl. Phys. B{\bf ~#1}, #2~(#3)}
\newcommand\CQG[3]{~Class. Quant. Grav.{\bf ~#1}, #2~(#3)}
\newcommand\PLB[3]{~Phys. Lett. B{\bf ~#1}, #2~(#3)}
\newcommand\PRL[3]{~Phys. Rev. Lett.{\bf ~#1}, #2~(#3)}
\newcommand\PR[3]{~Phys. Rep.{\bf ~#1}, #2~(#3)}
\newcommand\PRD[3]{~Phys. Rev. D{\bf ~#1}, #2~(#3)}
\newcommand\RMP[3]{~Rev. Mod. Phys.{\bf ~#1}, #2~(#3)}
\newcommand\SJNP[3]{~Sov. J. Nucl. Phys.{\bf ~#1}, #2~(#3)}
\newcommand\ZPC[3]{~Z. Phys. C{\bf ~#1}, #2~(#3)}
 \newcommand\IJGMP[3]{~Int. J. Geom. Meth. Mod. Phys.{\bf ~#1}, #2~(#3)}
  \newcommand\GRG[3]{~Gen. Rel. Grav.{\bf ~#1}, #2~(#3)}

\end{document}